\documentclass[aps,pra,preprint,superscriptaddress,amsmath,amssymb,showpacs]{revtex4-2}
\usepackage[english]{babel}
\usepackage{amssymb,bm}
\usepackage{graphicx}
\usepackage{amsmath}
\usepackage{color}
\usepackage{comment}
\usepackage[colorlinks=true,citecolor=blue,urlcolor=blue]{hyperref}
\begin{document}
 \begin{titlepage}

\title{Quasiparticles and ion cooling by an electron beam in a strong magnetic field}
\author{D.~M.~Popov}\email{D.M.Popov@inp.nsk.su}
\author{A.~I.~Milstein}\email{A.I.Milstein@inp.nsk.su}
\affiliation{Budker Institute of Nuclear Physics of SB RAS, 630090 Novosibirsk, Russia}

\date{\today}

\begin{abstract}
	The frictional force arising from the interaction of ions with a magnetized electron beam in a strong magnetic field is investigated. The problem is reduced to  the interaction of quasiparticles, or Larmor circles, with ions. A Hamiltonian describing this interaction is derived. The role of various impact parameters in the process of ion cooling is elucidated. The cases of both positively and negatively charged ions are considered.
	\end{abstract}

\maketitle
 \end{titlepage}

\section{Introduction}
Sixty years have passed since G.I.~Budker formulated the idea of electron cooling
of ions \cite{Budker1967}. The idea was successfully implemented, and currently numerous storage rings using the electron cooling method are operating throughout the world. Great progress has also been achieved in understanding the interaction of a cold electron beam with ions, and reviews devoted to this issue are regularly published \cite{PS1991,UFN2000,UFN2018,UFN2025}. Nevertheless, some important issues related to the interaction of magnetized electron beams with ions have not been fully understood. Even for the contribution to the friction force of the so-called Coulomb logarithm, obtained using perturbation theory, there are different expressions \cite{DS1978,Par1984}. It has been suggested \cite{Dik1988} that the reasons for the discrepancy in the results are related to the applicability range of perturbation theory. Furthermore, the contribution of the Coulomb logarithm is not large enough.
Therefore, to describe the electron cooling of ions in a strong magnetic field it is necessary to go beyond perturbation theory. Calculating the frictional force during the interaction of a magnetized electron beam with ions is a very complex task, even using modern computers. This approach is comparable to calculating the properties of condensed matter by accurately accounting for the interactions of electrons with each other and with atomic nuclei. L.D. Landau's idea of quasiparticles radically simplified theoretical research. It is this idea that forms the basis of modern condensed matter theory and nuclear physics (see, for example, \cite{LP1980}).

In our work, we follow the idea of L.D.~Landau and replace the interaction 
of ions with electrons moving in a strong magnetic field on the interaction of ions with the corresponding quasiparticles. As will be shown, these quasiparticles are macroscopic objects, namely, Larmor circles. We have calculated the effective Hamiltonian for the interaction of Larmor circles with ions.
As a result, the complex three-dimensional problem is reduced to a significantly simpler two-dimensional problem, allowing us to calculate the friction force outside the framework of perturbation theory. The cases of both positively and negatively charged ions are analyzed. The difference between the corresponding friction forces is entirely determined by the region of impact parameters in which perturbation theory fails.

\section{Interaction of electrons with ions in a strong magnetic field}
Let an ion move with velocity $\bm v^{(i)}$ in a strong magnetic field $\bm B$ directed along the $z$-axis. The ion interacts with a magnetized electron beam, and its interaction with the magnetic field can be neglected. We assume that the electron velocities along the magnetic field are much less than $v_i$. We direct the $y$-axis along the vector $\bm v^{(i)}_{\perp}$, which is the component of the vector $\bm v^{(i)}$ perpendicular to $\bm B$, and the $x$-axis along the vector $[\bm v^{(i)}\times\bm B]$. For convenience, we  assume that $\bm v^{(i)}_\parallel=\bm v^{(i)}-\bm v^{(i)}_\perp$ is antiparallel to $\bm B$. According to Newton's second law, the equation describing the motion of an electron interacting with a magnetic field and an ion has the form
\begin{align}\label{eq0}
	&\ddot{\bm r}_e=-\sigma\dfrac{|q_eq_i|}{m_e}\,\dfrac{\bm r_e-\bm r_i}{|\bm r_e-\bm r_i|^3}+[\bm\omega\times\dot{\bm r}_e]\,,
\end{align} 
where $\bm r_e$ and $\bm r_i$ are the radius vectors of the electron and ion, $q_e$ and $q_i$ are the charges of the electron and ion, $m_e$ is the mass of the electron, $\bm\omega=-q_e\bm B/(m_ec)$, $c$ is the speed of light, $\dot{\bm r_e}=\partial\bm r_e/\partial t$, $\sigma=\mbox{sgn}(-q_eq_i)$. The change in the velocity of the ion during interaction with the electron can be neglected, that is, $\dot{\bm r_i}=\bm v^{(i)}$. Let's move from the variable $\bm r_e$ to the variable $\bm \xi=\bm r_e-\bm r_i$, then
\begin{align}\label{eq1}
	&\ddot{\bm\xi}=-\sigma\dfrac{|q_eq_i|}{m_e}\,\dfrac{\bm \xi}{\xi^3}+[\bm\omega\times(\dot{\bm \xi}+\bm v^{(i)}_\perp)]\,,
\end{align} 
Let us now introduce, instead of the variables $t$ and $\bm \xi$, the dimensionless variables $\tau$ and $\bm R$,
 $$\tau=\omega t\,,\quad \bm R=\bm\xi/\lambda\,,\quad   \lambda=\left(\dfrac{|q_eq_i|}{m_e\omega^2}\right)^{1/3}\,.$$
As a result we obtain 
\begin{align}\label{eq2}
	&{\bm R}''=-\sigma\dfrac{\bm R}{R^3}+[\bm\nu\times({\bm R}'+\bm \beta_\perp)]\,,
\end{align}
where $\bm\nu=\bm B/B$, $\bm\beta_\perp=\bm v^{(i)}_\perp/(\omega\lambda)$ and ${\bm R}'=\partial\bm R/\partial \tau$. In Cartesian coordinates, $\bm R=(X,Y,Z)$, we have a system of equations
\begin{align}\label{eq3}
	&{X}''=-\sigma\dfrac{X}{R^3}-(Y'+\beta_\perp)\,,\nonumber\\
	&{Y}''=-\sigma\dfrac{Y}{R^3}+X'\,,\nonumber\\
	&{Z}''=-\sigma\dfrac{Z}{R^3}\,,\quad R=(X^2+Y^2+Z^2)^{1/2}\,.
\end{align}	

Let's represent the  motion of electron as the motion of the center of the Larmor circle and its rotation around this center. To do this, we introduce five variables: $X_c$, $Y_c$, $Z_c$, $\rho$, and $\phi$, as well as two constraint equations between them,
\begin{align}\label{def}
	&{X}=X_c+\rho\cos\phi\,,\quad Y=Y_c+\rho\sin\phi\,,\quad Z=Z_c\,,\nonumber\\
	&{X}'=-\rho\sin\phi\,,\quad Y'=\rho\cos\phi-\beta_\perp\,.
\end{align}	
Here $\rho$ can be interpreted as the radius of the Larmor circle, $\phi$ is the azimuth angle describing the rotation of  electron in the circle, and $\bm R_c=(X_c,Y_c,Z_c)$ is the vector between the point of ion  location and the center of the Larmor circle.

After simple transformations, we find equations for the new variables
\begin{align}\label{eqnew}
	&{\rho}'=\sigma\dfrac{X_c\sin\phi-Y_c\cos\phi}{R^3}\,,\quad
	{\phi}'=\sigma\dfrac{\rho +X_c\cos\phi+Y_c\sin\phi}{\rho\,R^3}+1\,,\nonumber\\
	&{X_c}'=\sigma\dfrac{Y_c+\rho\sin\phi}{R^3}\,,\quad {Y_c}'=-\sigma\dfrac{X_c+\rho\cos\phi}{R^3}-\beta_\perp\,,\quad
	{Z_c}''=-\sigma\dfrac{Z_c}{R^3}\,,\nonumber\\ &R=[R_c^2+\rho^2+2\rho(X_c\cos\phi+Y_c\sin\phi)]^{1/2}\,.
\end{align}	
Now, instead of a single electron, we consider a system of electrons with a random phase $\phi$ at large negative times. The central point in the transition from the interaction of ions with an electron beam to the interaction with quasiparticles is the assumption of a random phase $\phi$ at all times. Based on this, we average the equations of motion for $\bm R_c$ and $\rho$ in \eqref{eqnew} over $\phi$. As a result, we obtain
\begin{align}\label{eqnew1}
	&\bar{\rho}'=0\,,\quad
	\bar{X_c}'=\sigma\dfrac{\mathcal F}{\bar R^3}\,\bar Y_c\,,\quad \bar{Y_c}'=-\sigma\dfrac{\mathcal F}{\bar R^3}\,\bar X_c-\beta_\perp\,,\quad \bar{Z_c}''=-\sigma\dfrac{\mathcal F_0}{\bar R^3}\,\bar Z_c\nonumber\\
	&\mathcal F=\left(1-\dfrac{\bar R^2}{2\bar R_\perp^2}\right)s^{3/2}P_{1/2}(s)+\dfrac{\bar R^2}{2\bar R_\perp^2}s^{1/2}P_{-1/2}(s)\,,\quad 	\mathcal F_0=s^{3/2}P_{1/2}(s)\,,\nonumber\\ &\bar R^2=\bar R_\perp^2+\bar{Z_c}^2+\bar \rho^2\,,\quad s=\dfrac{{\bar R}^2}{\sqrt{\bar R^4-4\bar \rho^2\bar R_\perp^2}}\,,\quad \bar R_\perp^2=\bar{X_c}^2+\bar{Y_c}^2\,.
\end{align}	
Here $P_\nu(s)$ is the Legendre function. It is clear that in our approach $\bar\rho$ is independent of time, so that it is an integral of motion. The initial conditions for the system \eqref{eqnew1} at large absolute values and negative $\tau$ are
 \begin{align}\label{initial}
	& 	\bar{X_c}'(\tau)=0\,,\quad \bar{Y_c}'(\tau)=-\beta_\perp\,,\quad \bar{Z_c}'(\tau)=\beta_\parallel\,, \nonumber\\
	&\bar{X_c}(\tau)=b_x\,,\quad \bar{Y_c}(\tau)=b_y-\beta_\perp\,\tau,\quad \bar{Z_c}(\tau)=\beta_\parallel \tau\,,
\end{align}	
   where $\beta_\parallel=v^{(i)}_\parallel/(\omega\lambda)$, the vector $\bm b=(b_x,b_y,0)$ is a dimensionless impact parameter.
   It is essential that the system of equations \eqref{eqnew1} corresponds to the Hamiltonian
   \begin{align}\label{H}
   	&H= \dfrac{1}{2}{\cal P}^2_z-\dfrac{\sigma}{\bar R}\,s^{1/2}P_{-1/2}(s)+\beta_\perp \bar X_c\,,\nonumber\\
   	&\bar R^2=h_\perp+\bar{Z_c}^2+\bar \rho^2\,,\quad s=\dfrac{{\bar R}^2}{\sqrt{\bar R^4-4\bar \rho^2h_\perp}}\,,\quad h_\perp=\bar{X_c}^2+{\cal P}_x^2\,.
   \end{align}	
 Note that $s\geq 1$ and $P_{-1/2}(s)>0$.
 Thus, we have reduced the problem of three-dimensional motion in the space $\bar X_c\bar Y_c\bar Z_c$ to the problem of two-dimensional motion in the plane $\bar X_c\bar Z_c$. In the latter case, the canonical momentum ${\cal P}_x$ conjugate to the coordinate $\bar X_c$ is $\bar Y_c$ of the original problem, $\bar Y_c={\cal P}_x$.  

Using Eqs.~\eqref{eq1}, \eqref{eqnew1} and the  momentum conservation law,
we find the components of the friction force $\bm F$ acting on the ion,
\begin{align}\label{FXYZ}
	&F_x=0\,,\quad F_y=N\,{\cal F}_y\,,\quad
	F_z=N\,{\cal F}_z\,, \nonumber\\
	& {\cal F}_y= {\beta_\parallel}\int d^2b\,\Delta \bar X_c(\bm b)\,,\quad
	{\cal F}_z=-{\beta_\parallel}\int d^2b\,\Delta {\cal P}_z(\bm b)\,,\quad N=n_e\lambda\,|q_eq_i|\,,
\end{align}	
where $n_e$ is the electron density in the beam, $\Delta \bar X_c(\bm b)$ is the total change in the coordinate $\bar X_c(\bm b)$ during scattering, $\Delta {\cal P}_z(\bm b)$ is the total change in the momentum ${\cal P}_z$ during scattering. The quantities $\Delta {\cal P}_z(\bm b)$ and $\Delta \bar X_c(\bm b)$ are not independent. Indeed, from the energy conservation law corresponding to the Hamiltonian $H$ \eqref {H} and the infinity motion, the following relation holds
\begin{align}\label{rel}
	& \dfrac{1}{2}\Big(\Delta {\cal P}_z(\bm b)\Big)^2+\beta_\parallel \Delta {\cal P}_z(\bm b)+\beta_\perp\,\Delta \bar X_c(\bm b)=0\,.
\end{align}	
Since $\bar Y_c={\cal P}_x$, then integration over $d^2b$ in Eq.~\eqref{FXYZ}
is equivalent to integration over $d\bar X_c\,d{\cal P}_x$, that is, over the initial phase space of the electron beam.

The relation \eqref{rel} has an important consequence for $W$, which is the change rate of the ion energy,
\begin{align}\label{W}
	&W=\bm F\cdot\bm v=\lambda\omega(\beta_\perp F_y-\beta_\parallel F_z)
	=
	\lambda\omega\beta_\parallel N\int d^2b\,[\beta_\perp\Delta \bar X_c(\bm b)+\beta_\parallel \Delta {\cal P}_z(\bm b)] \nonumber\\
	&	=- \dfrac{1}{2}\lambda\omega\beta_\parallel N\int d^2b\,\Big(\Delta {\cal P}_z(\bm b)\Big)^2\,.
\end{align}	
Thus, $W$ is always negative, which is natural for friction, although the force component $F_y$ can be both positive and negative.
The sign of $F_y$ affects the change in the ion beam distribution function over the direction of $\bm v^{(i)}$.

To understand the physical content of the model, it is useful to start from the  consideration of several special cases.

\section{Limit $\bar\rho\to 0 $}

The limit $\bar\rho\to 0 $ corresponds to the case of small size of Larmor circles compared to the impact parameters. Note that the dimensionless parameter $\bar\rho$ is determined by the transverse electron velocity $v^{(e)}_\perp$  and the magnetic field strength, $\bar\rho\sim v^{(e)}_\perp/(\omega\lambda)$, and can be either greater or less than unity.

When $\bar\rho\to 0$ we have $s=1$, $P_{\pm 1/2}(1)=1$, and Eq.~\eqref{eqnew1} goes over to 
\begin{align}\label{rho0}
	&	\bar{X_c}'=\sigma\dfrac{\bar Y_c}{\bar R^3}\,\,,\quad \bar{Y_c}'=-\sigma\dfrac{\bar X_c}{\bar R^3}-\beta_\perp\,,\quad \bar{Z_c}''=-\sigma\dfrac{\bar Z_c}{\bar R^3}\,,\nonumber\\
	&\bar R^2=\bar{X_c}^2+\bar{Y_c}^2+\bar{Z_c}^2\,.
\end{align}	
The initial conditions for this system coincide with Eq.~\eqref{initial}. It is important that when integrating over the impact parameters in Eq.~\eqref{FXYZ} no divergence occurs at $b \to 0$.
 This is due to the fact that the integrals in \eqref{FXYZ} are not invariant under scaling with respect to $b$. That is, the lower limit of $b_{\min}$ in the Coulomb logarithm $\log(b_{\max}/b_{\min})$ is not $b_{\min}\sim\bar\rho$, but $b_{\min}\sim 1/\beta^2$. The latter condition corresponds to  the applicability of perturbation theory. The parameter $b_{\max}$ can be estimated as $b_{\max}\sim n^{-1/3}/\lambda\gg 1/\beta^2$ \cite{Dik1988,Meshkov1994}.
 
 \subsection{Case $\beta_\perp=0$}
 When $\beta_\perp=0$ and $\bar\rho=0$, we find that $R_\perp^2=\bar{X_c}^2+\bar{Y_c}^2$  is independent of time,  $R_\perp^2=b_x^2+b_y^2=b^2$. Therefore, $\bar{X_c}=b\cos\psi$, $\bar{Y_c}=b\sin\psi$. For the phase $\psi$, we have 
 \begin{align}\label{psi}
 	&	\psi'=-\dfrac{\sigma}{(\bar Z_c^2+b^2)^{3/2}}\,.
 \end{align}	
It  follows from this equation that 
 \begin{align}\label{DelX}
 	&\Delta \bar X_c(\bm b)=b\,[\cos(\psi_0+\Delta\psi)-\cos\psi_0]\,,\nonumber\\
 	&\Delta\psi=-\int_{-\infty}^\infty \dfrac{\sigma\,d\tau}{[\bar Z_c^2(\tau)+b^2]^{3/2}}\,.
 \end{align}	
Since $\Delta\psi$ is independent of $\psi_0$ and $d^2b=b\,db\,d\psi_0$, then after integrating over $\psi_0$ in Eq.~\eqref{FXYZ} we obtain $F_y=0$, which is obvious from the reasons of symmetry  at $\beta_\perp=0$.

The energy conservation law  describing one-dimensional motion along the $z$ axis has the form
\begin{align}\label{Ez}
	&	\dfrac{\beta_\parallel^2}{2}=\dfrac{\bar Z_c'^2}{2}-\dfrac{\sigma}{\sqrt{\bar Z_c^2+b^2}}\,.
\end{align}	
Thus, we have one-dimensional motion with positive energy in the potential $U(z)=-\sigma/\sqrt{z^2+b^2}$. Consequently, for $\sigma=1$ we have $\Delta {\cal P}_z(\bm b)=0$ and $F_z=0$ for any $b$. For $\sigma=-1$ we have $\Delta {\cal P}_z(\bm b)=0$ for
$\beta_\parallel^2/2>1/b$ and $\Delta {\cal P}_z(\bm b)=-2\beta_\parallel$ for $\beta_\parallel^2/2<1/b$. As a result, for $\sigma=-1$ we find
\begin{align}\label{Fzminus}
	&F_z=N\dfrac{8\pi}{\beta^2}\,.
\end{align}	
The difference in frictional force for positively and negatively charged ions is evident. This difference is determined by the region of impact parameters for which perturbation theory for electron motion is not applicable.

We now turn to the case $\beta_\perp\neq 0$.
\subsection{Perturbation theory}
For perturbation theory to be applicable, it is necessary that $|\Delta {\cal P}_z|\ll \beta_\parallel$. When integrating over $b$ we introduce an upper limit of integration $b_{\max}$. To clarify the influence of small impact parameters $b\sim\bar\rho$, we replace $1/R$ with $1/\sqrt{R^2+\bar\rho^2}$ in the Coulomb potential. Using perturbation theory, after direct but cumbersome calculations, we obtain the logarithmic contribution to the frictional force, 
\begin{align}\label{PT}
	&F_z=N\dfrac{4\pi\beta_\parallel\beta_\perp^2}{\beta^5}\,\int_{b_{\min}}^{b_{\max}}
	\Bigg[b^2+\bar\rho^2\,\log\dfrac{b_{\max}}{b}\Bigg]\,\dfrac{b\,db}{(b^2+\bar\rho^2)^2}\,,\nonumber\\
	&F_y=N\dfrac{2\pi\beta_\perp}{\beta^5}\,\int_{b_{\min}}^{b_{\max}}
	\Bigg[(\beta_\parallel^2-\beta_\perp^2)b^2+2\bar\rho^2\beta_\parallel^2\,\log\dfrac{b_{\max}}{b}\Bigg]\,\dfrac{b\,db}{(b^2+\bar\rho^2)^2}\,.	
\end{align}	
where $b_{\min}=2/\beta^2$, the factor $\log(b_{\max}/b)$ in the integrand arises as a result of integration over time. The condition $b\gtrsim b_{\min}=2/\beta^2$ follows from the criterion of the applicability of perturbation theory. If $\bar\rho\ll b_{\min}$, then the second term in the square brackets in the integrand is small, and we arrive at the result obtained in \cite{Par1984},
\begin{align}\label{par}
	&F_z=N\dfrac{4\pi\beta_\parallel\beta_\perp^2}{\beta^5}\,\log\dfrac{b_{\max}}{b_{\min}}\,,\nonumber\\
	&F_y=N\dfrac{2\pi\beta_\perp}{\beta^5}(\beta_\parallel^2-\beta_\perp^2)\,\log\dfrac{b_{\max}}{b_{\min}}\,.
\end{align}	
For arbitrary $\bar\rho\ll b_{\max}$, with logarithmic accuracy we obtain
\begin{align}\label{DS}
	&F_z=N\dfrac{2\pi\beta_\parallel\beta_\perp^2}{\beta^5}\Bigg[2+\dfrac{\bar\rho^2}{\bar\rho^2+b_{\min}^2}\Bigg]\,\log\left(\dfrac{b_{\max}}{\bar\rho+b_{\min}}\right)\,,\nonumber\\
	&F_y=N\dfrac{2\pi\beta_\perp}{\beta^5}\Bigg[(\beta_\parallel^2-\beta_\perp^2)+\beta_\parallel^2\dfrac{\bar\rho^2}{\bar\rho^2+b_{\min}^2}\Bigg]\,\log\left(\dfrac{b_{\max}}{\bar\rho+b_{\min}}\right)\,.
\end{align}	
For $\bar\rho\gg b_{\min}$, this result coincides with that of \cite{DS1978}. Consequently, the results of \cite{DS1978} and \cite{Par1984} have different ranges of applicability, $\bar\rho\gg b_{\min}$ and $\bar\rho\ll b_{\min}$, respectively. As expected, the logarithmic contribution to the friction force \eqref{DS} is independent of  $\sigma=\pm 1$, since it is obtained using perturbation theory. Therefore, the difference between the friction force for positively and negatively charged ions is determined solely by the region of impact parameters for which perturbation theory is not applicable. It is very important that our model of the quasiparticle-ion interaction, based on the use of the Hamiltonian \eqref{H}, does not require the introduction of any regularization parameter $b_{\min}$ at small distances, since it does not contain any divergences as $b\to 0$. It is applicable for all $b\ll b_{max}$ with $b_{\max} \gg 1$.

From Eq.~\eqref{DS} it is evident that the friction force $\bm F$ depends strongly on the ratio $\beta_\parallel/\beta_\perp$. The component $F_z$ is always positive
while $F_y$ changes its sign depending on $\beta_\parallel/\beta_\perp$, and the sign change occurs at different values of this quantity depending on $\bar\rho$.
Note that  the expression for  $W=\bm F\cdot\bm v=\lambda\omega(\beta_\perp F_y-\beta_\parallel F_z)$   has the form
 \begin{align}\label{Was}
 	&W=-\lambda\omega N\dfrac{2\pi\beta^2_\perp}{\beta^3}\,\log\left(\dfrac{b_{\max}}{\bar\rho+b_{\min}}\right)
 \end{align}	
 for any ratio between $\bar\rho$ and $b_{\min}$, as  follows from Eq.~\eqref{DS}.

\section{Dynamics of quasiparticle-ion interaction for any $\bar\rho$}
Let us consider  parameters  typical for all electron cooling setups \cite{UFN2025, Dik1988, ESR1997}.
Magnetic field $B=4\cdot 10^3\,\mbox {G}$, corresponding
frequency $\omega\sim 7\cdot 10^{10}\,\mbox{s}^{-1}$. For $H^{\pm}$ ions, $\lambda=(m_ec^2/B^2)^{1/3}=4\cdot 10^{-5}\,\mbox{cm}$.
For heavy ions, $\lambda$ can be approximately 4 times greater. The typical spread of ion velocities is $ v^{(i)}\sim 3\cdot 10^{6}\,\mbox {cm/s}$. Then  $\beta\sim 1$ for $H^{\pm}$ and $\beta\sim 0.2$ for heavy ions. The spread of electron transverse velocities is  $v^{(e)}_\perp\sim 3\cdot 10^7\,\mbox {cm/s}$. Thus, $\bar\rho\sim 5$ for $H^{\pm}$ and $\bar\rho\sim 1$ for heavy ions.
These parameters can vary significantly depending on the magnetic field strength and the temperature of the electron and ion beams.

The Coulomb potential $1/R$ changes at large distances due to electron beam screening effects. We take this effect into account by making the substitution
$$\dfrac{1}{R}\to \dfrac{1}{R}\,\exp{(-R/b_{\max})}\,.$$
The dimensionless quantity $b_{\max}$ can be estimated as follows \cite{Dik1988,Meshkov1994},
\begin{align}\label{bmax}
	b_{\max}\sim \max\left(\dfrac{n^{-1/3}}{\lambda},\, \dfrac{v^{(i)}}{\omega_p\lambda}\right)\,,\quad \dfrac{v^{(i)}}{\omega_p\lambda}=\dfrac{\omega}{\omega_p}\beta\,,
\end{align}	
where $\omega_p=\sqrt{4\pi n_ee^2/m_e}$ is the plasma frequency of electrons. For $n_e=10^9\,\mbox{cm}^{-3}$, we obtain $\omega_p\sim 1.7\cdot 10^9\,\mbox{s}^{-1}$ and $b_{\max}\sim 50$. The condition $b_{\max}\sim n^{-1/3}/\lambda$ in \eqref{bmax} corresponds to the fact that for $b<b_{\max}$ there must be many electrons. The condition $b_{\max}\sim v^{(i)}/(\omega_p\lambda)$ corresponds to the distance at which the electron matter reacts on a perturbation created by an ion moving with velocity $v^{(i)}$.

\begin{figure}[b]
	\centering
	\includegraphics[width=0.47\linewidth]{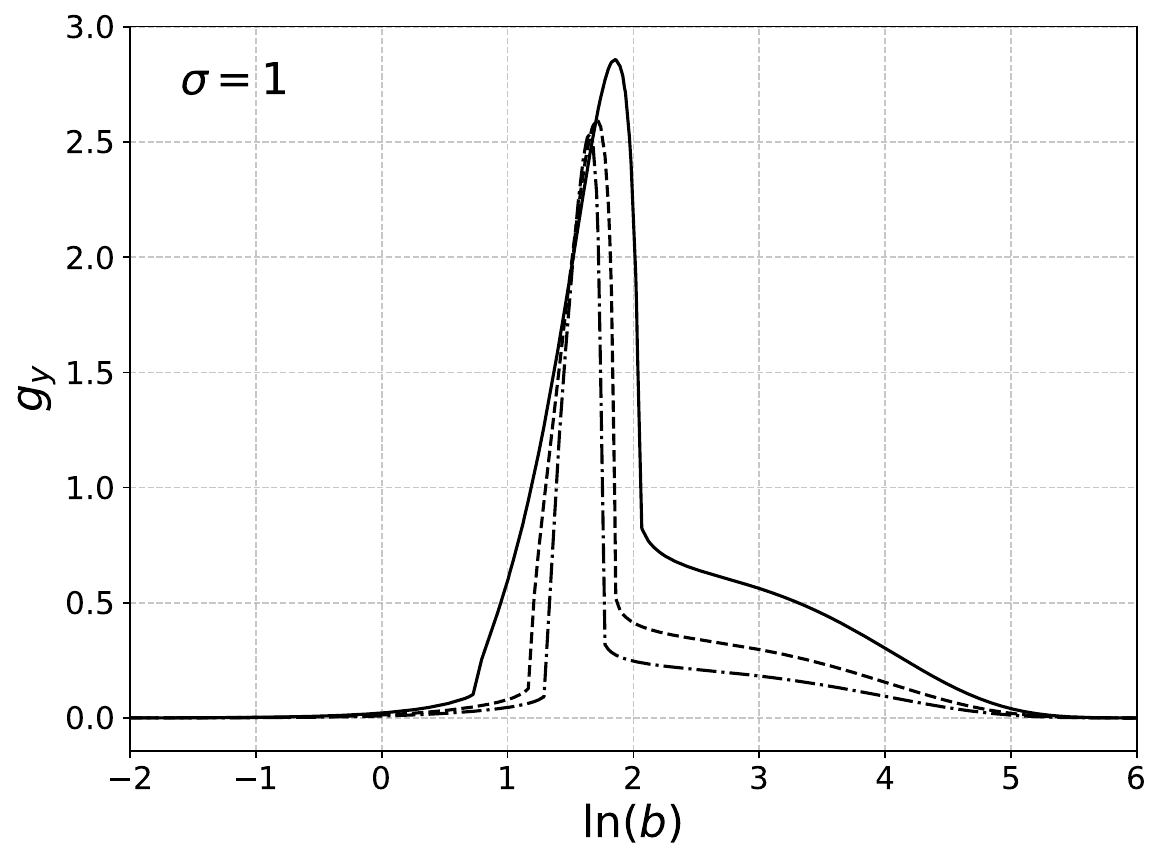}
	\includegraphics[width=0.47\linewidth]{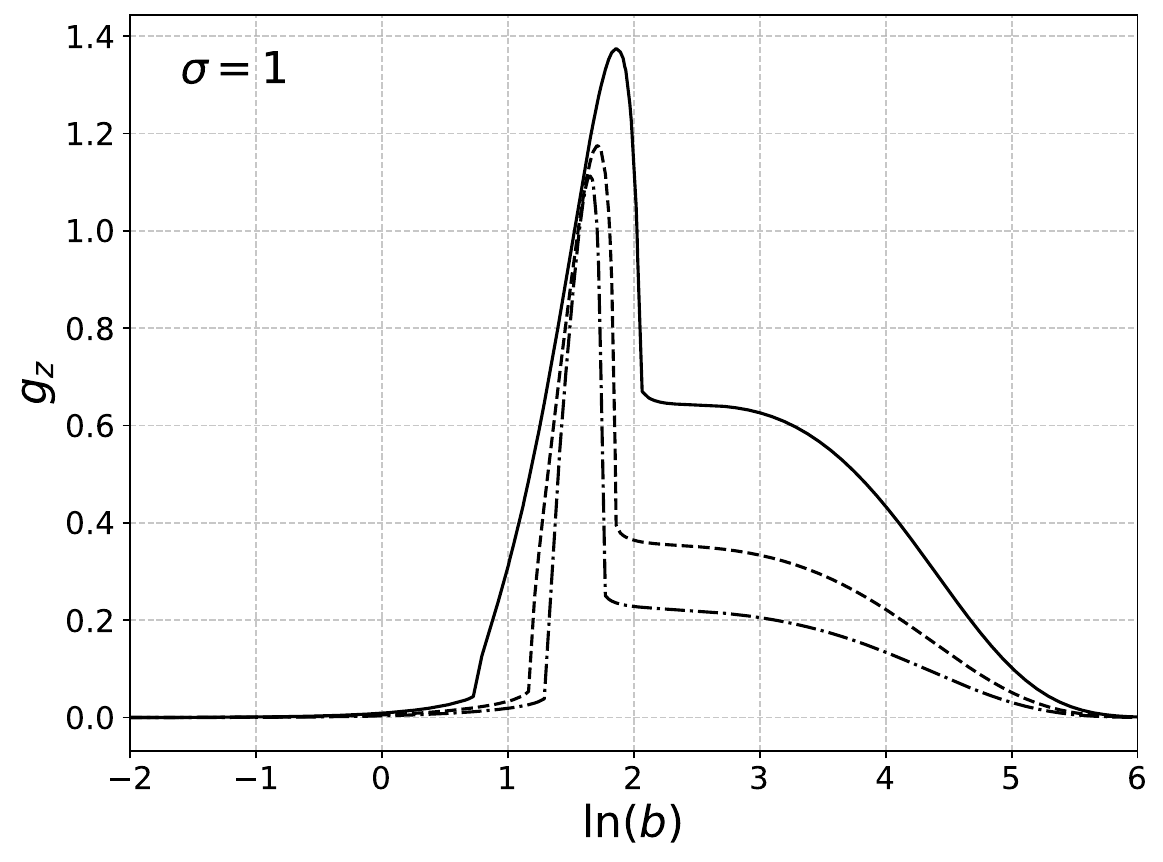}
	\includegraphics[width=0.47\linewidth]{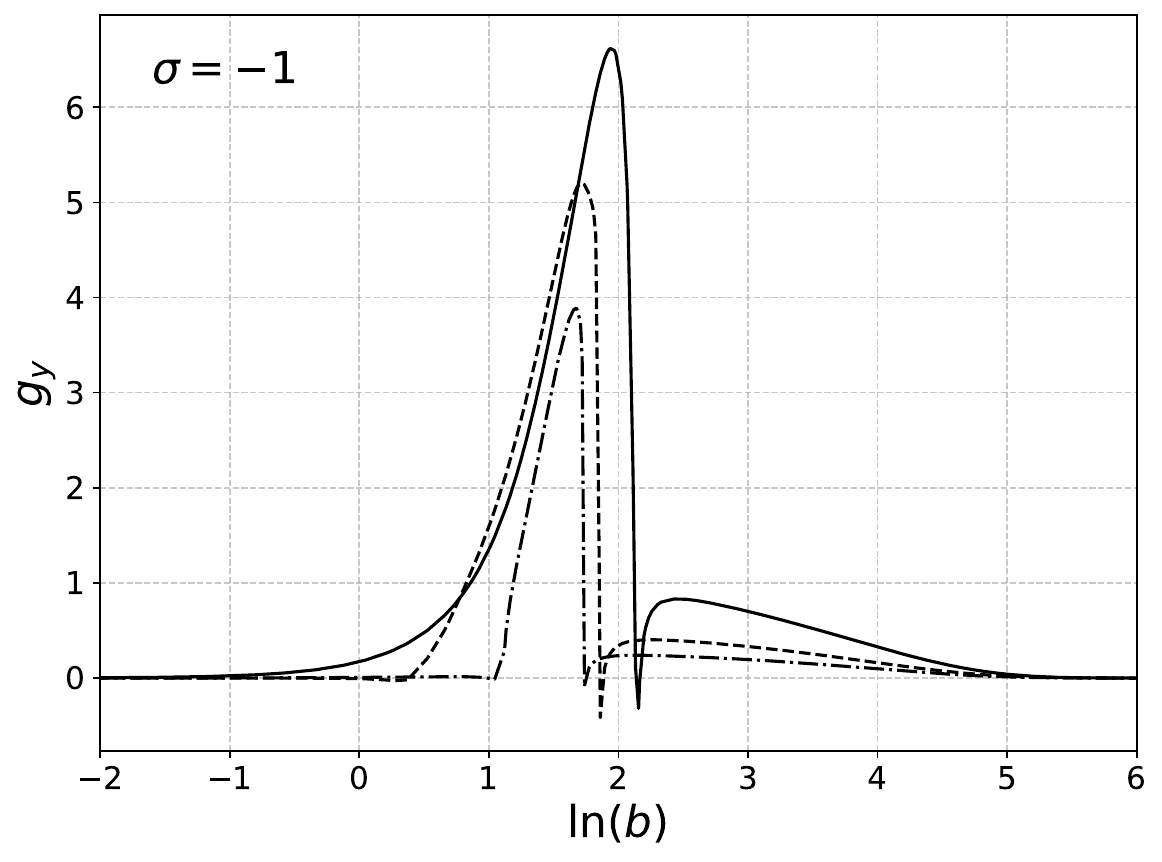}
	\includegraphics[width=0.47\linewidth]{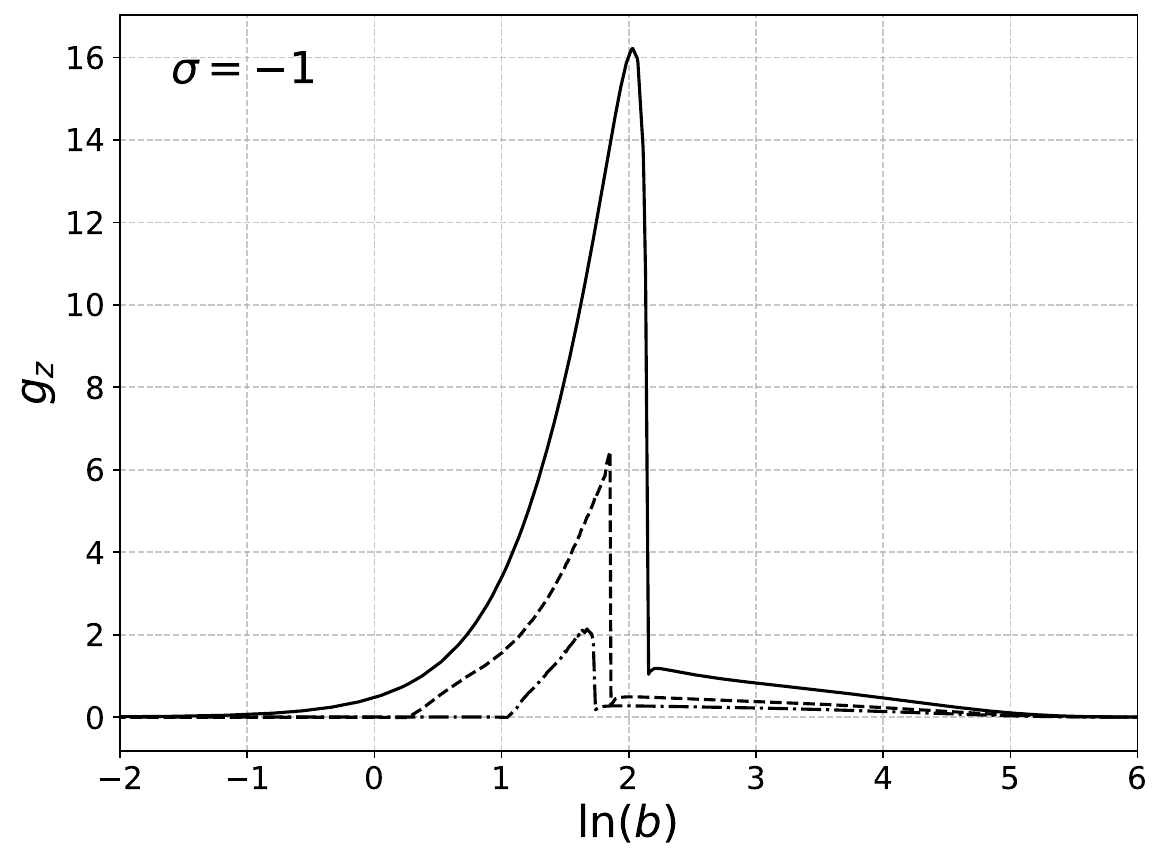}
	\caption{Dependence of $g_y(b)$ and $g_z(b)$ on $\log b$ for $\bar\rho=5$, $\theta=\pi/8$, $\sigma=\pm 1$, $b_{\max}=100$, $\beta=0.55$ (solid line), $\beta=0.78$ (dashed line), and $\beta=1$ (dash-dotted line).}
	\label{Fyzbeta}
\end{figure}

To understand the  interaction of quasiparticles with an ion outside the framework of perturbation theory, it is convenient to make the substitution of variables $b_x=b\cos\phi$ and $b_y=(b\beta/\beta_\parallel)\sin\phi$.
Fig. \ref {Fyzbeta} shows the dependence of
$$g_y(b)=\beta b^2\int_0^{2\pi}\Delta X(b,\phi)\dfrac{d\phi}{2\pi}\,,\quad g_z(b)=-\beta b^2\int_0^{2\pi}\Delta {\cal P}_z(b,\phi)\dfrac{d\phi}{2\pi}\,$$
on $\log b$ for $\bar\rho=5$, $\sigma=\pm 1$, $b_{\max}=100$, $\theta=\arcsin(\beta_\parallel/\beta)=\pi/8$, and several values of $\beta$.
Since $\int b\,g_{y,z}\,db=\int b^2g_{y,z}\,d\ln b$, the areas under the curves in Fig.\ref{Fyzbeta} coincide with the corresponding components of the friction force $F_{y,z}$ up to a factor. It is evident that for $\sigma=1$ (positively charged ions), the contribution to the force from the peak region is comparable to the logarithmic contribution determined by perturbation theory. The latter region is located to the right of the peak region in Fig.\ref{Fyzbeta}. For $\sigma=-1$ (negatively charged ions), the contribution to the friction force from the peak region significantly exceeds the logarithmic contribution. This observation emphasizes the fundamental importance of account for nonperturbative effects when calculating the friction force.

\begin{figure}[h!]
	\centering
	\includegraphics[width=0.47\linewidth]{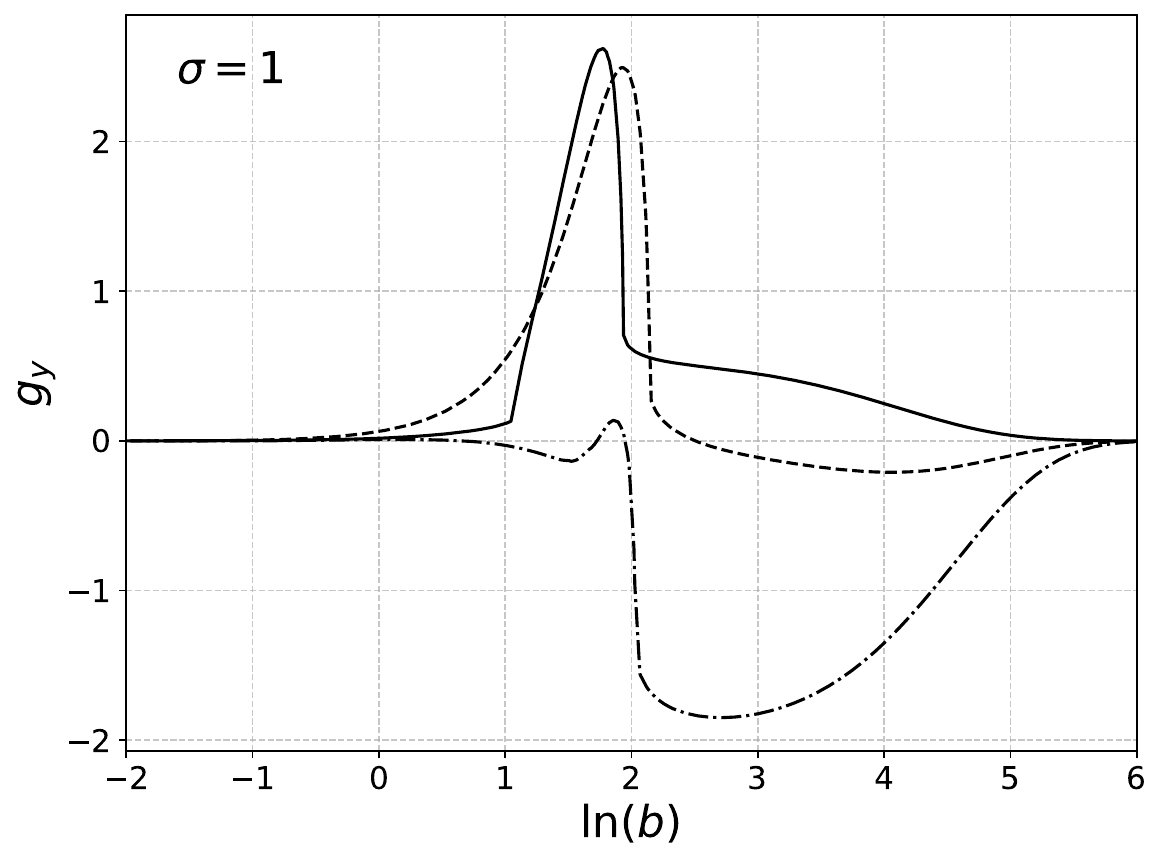}
	\includegraphics[width=0.47\linewidth]{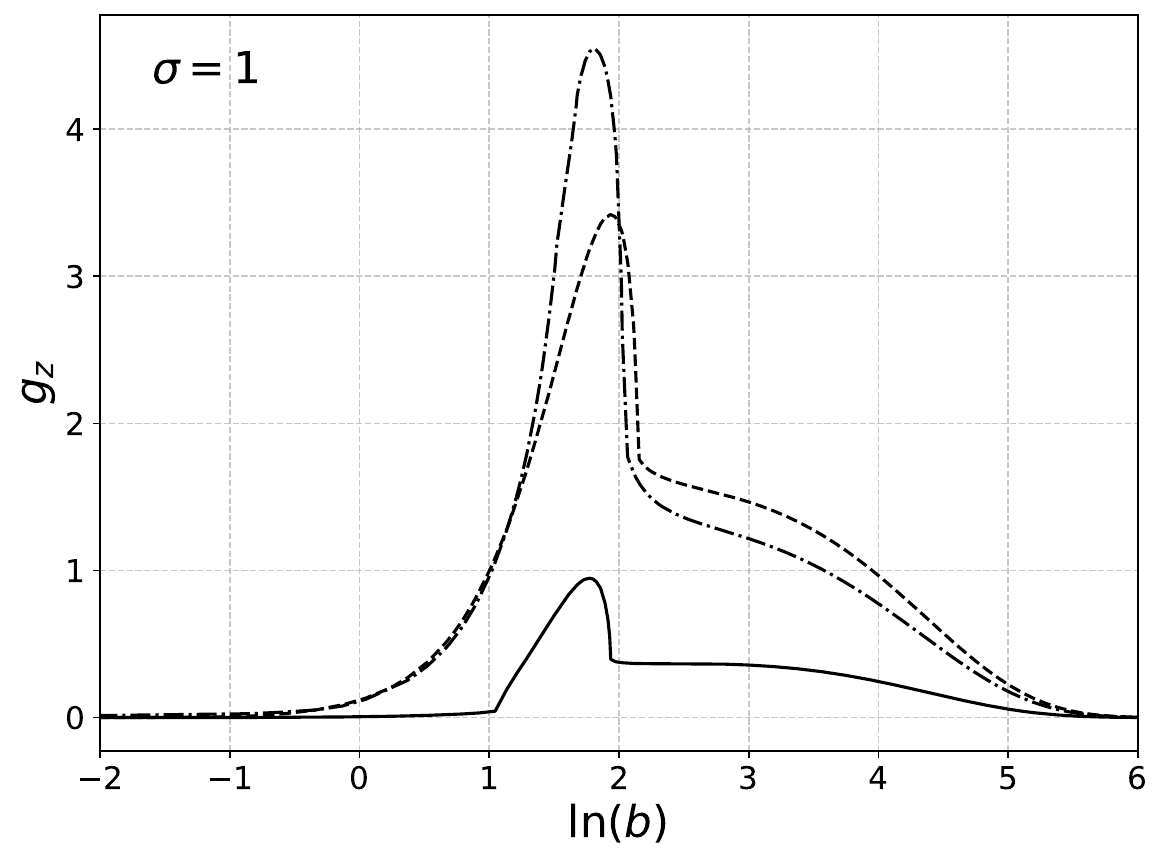}
	\includegraphics[width=0.47\linewidth]{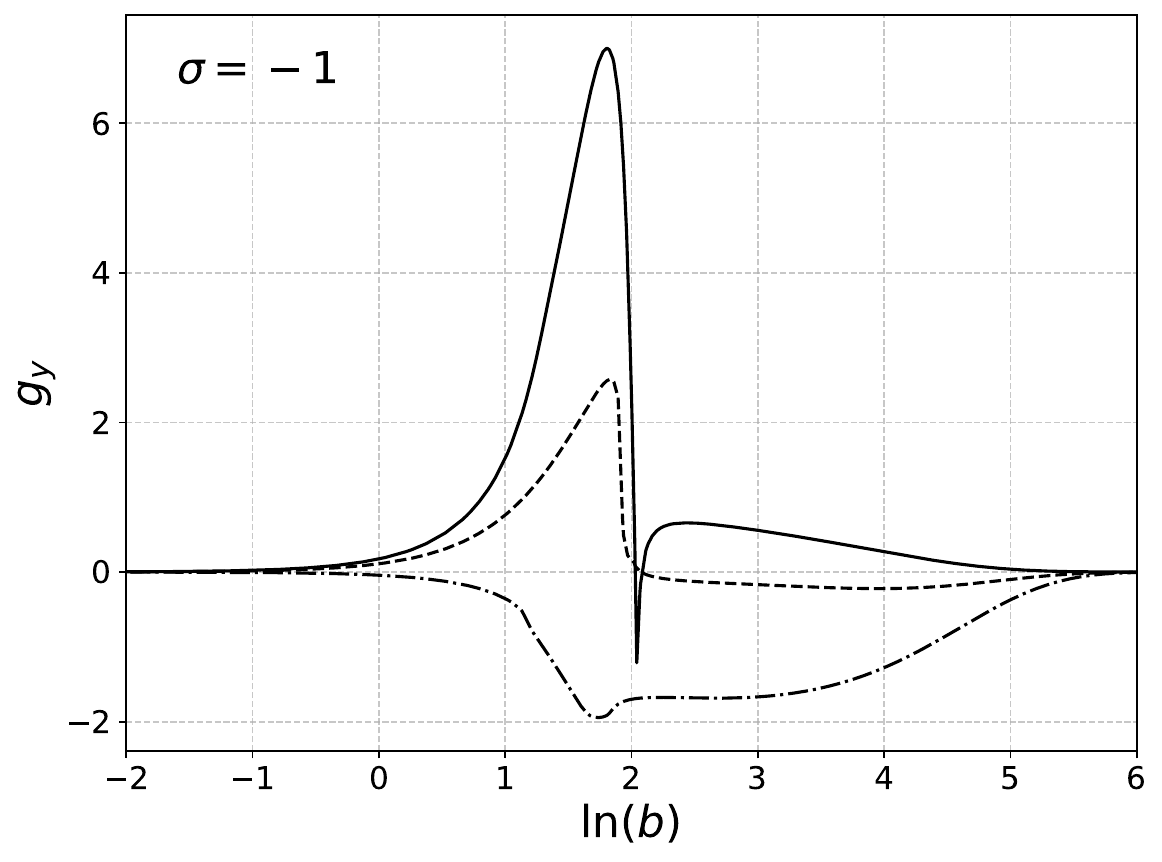}
	\includegraphics[width=0.47\linewidth]{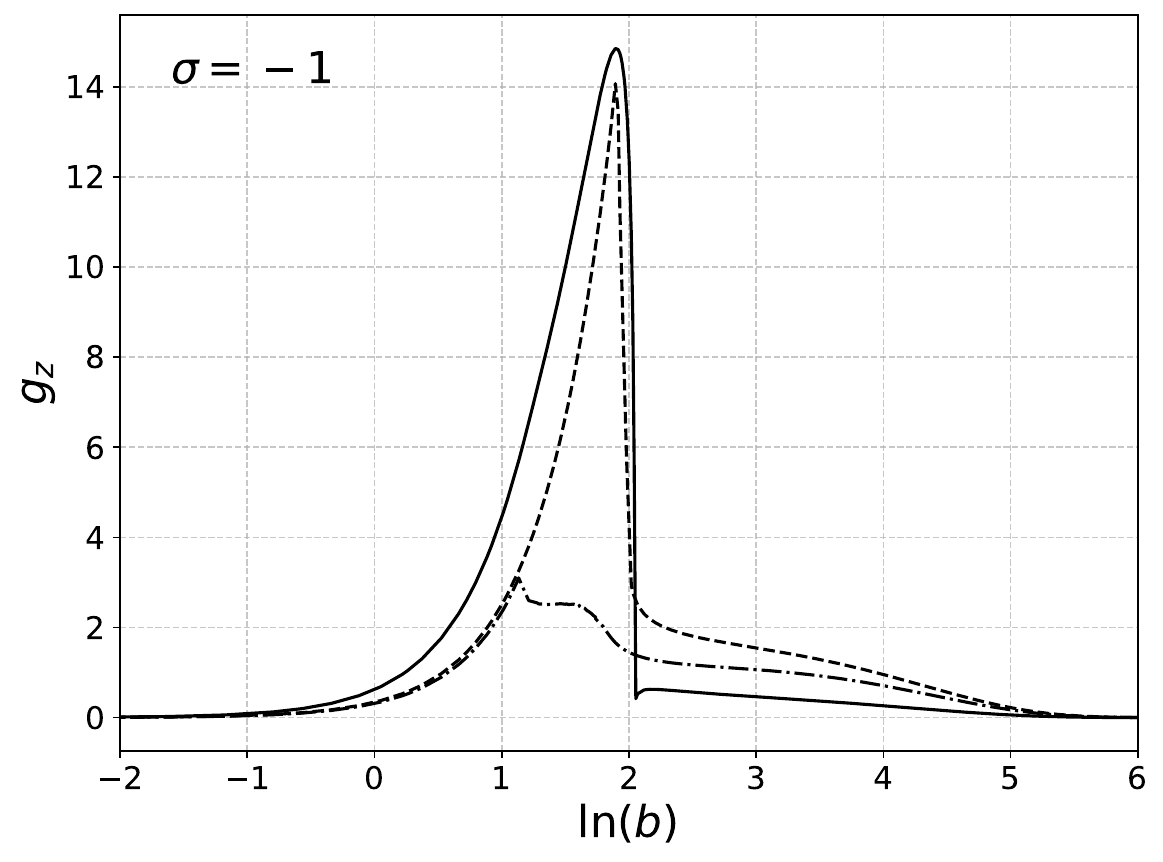}
	\caption{Dependence of $g_y(b)$ and $g_z(b)$ on $\log b$ for $\bar\rho =5$, $\beta=0.6$, $\sigma=\pm 1$, $b_{\max}=100$, $\theta=\pi/10$ (solid line), $\theta=\pi/4$ (dashed line) and $\theta=2\pi/5$ (dash-dotted line).}
	\label{Fyztheta}
\end{figure}

\begin{figure}[h!]
	\centering
	\includegraphics[width=0.47\linewidth]{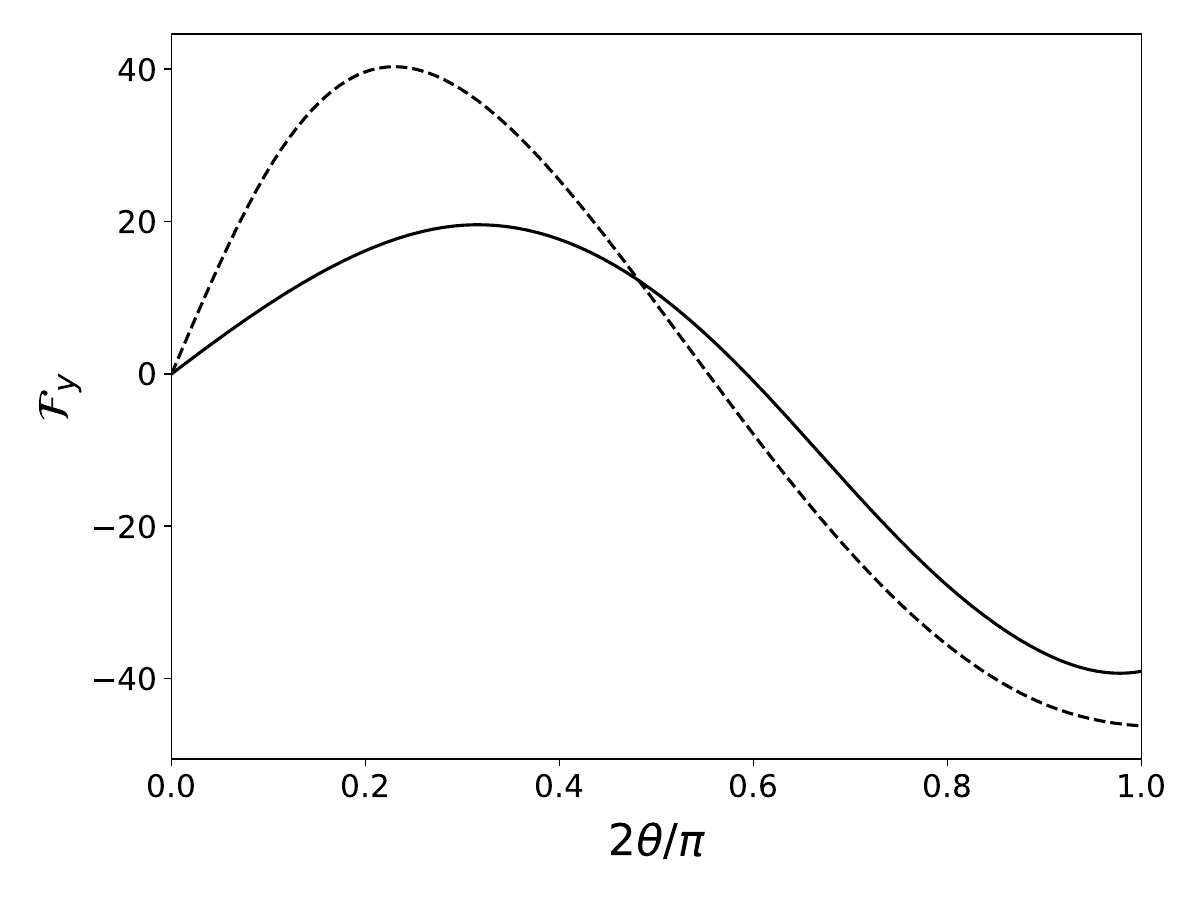}
	\includegraphics[width=0.47\linewidth]{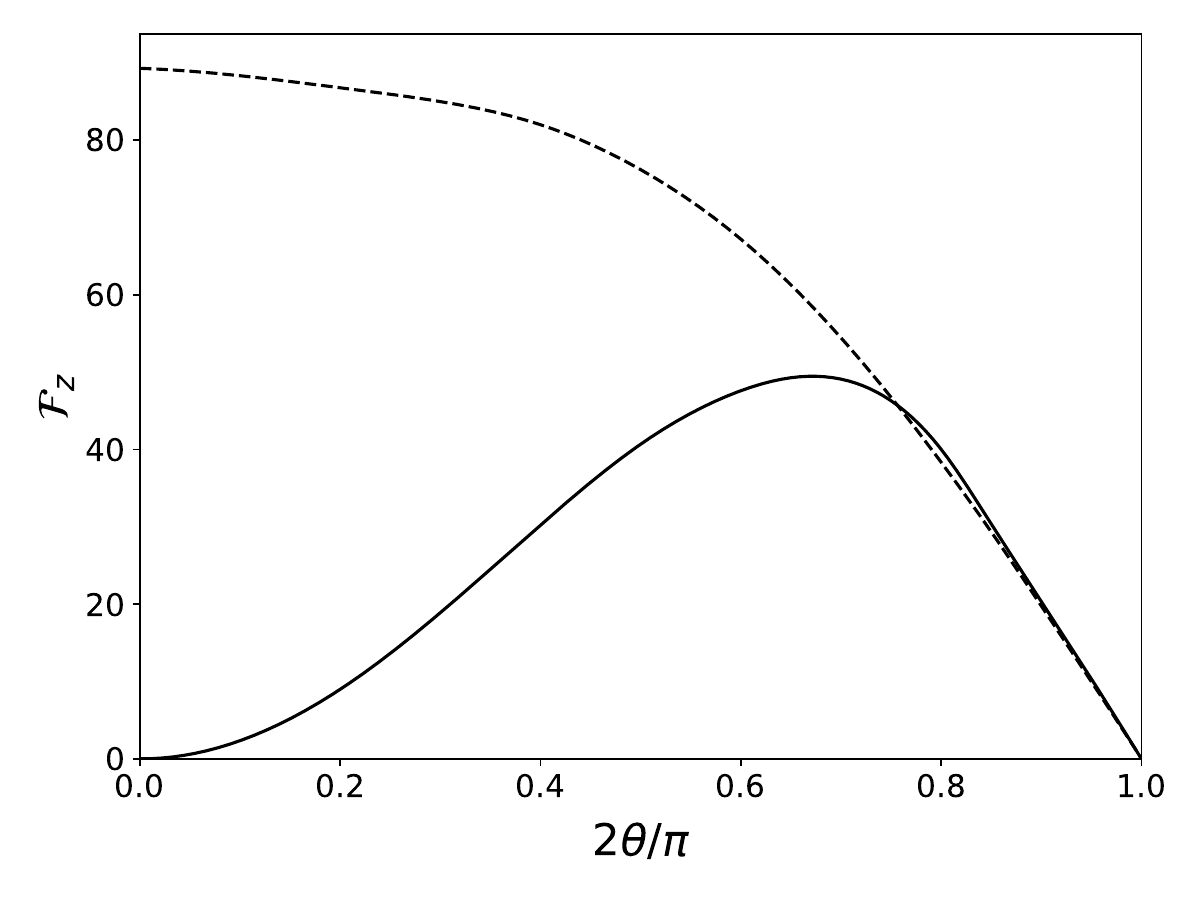}
	\caption{Dependence of ${\cal F}_y$ and ${\cal F}_z$ on $\theta$ for $\bar\rho =5$, $\beta=0.6$, $b_{\max}=100$, $\sigma=1$ (solid line) and $\sigma=-1$ (dashed line).}
	\label{totFyztheta}
	\includegraphics[width=0.47\linewidth]{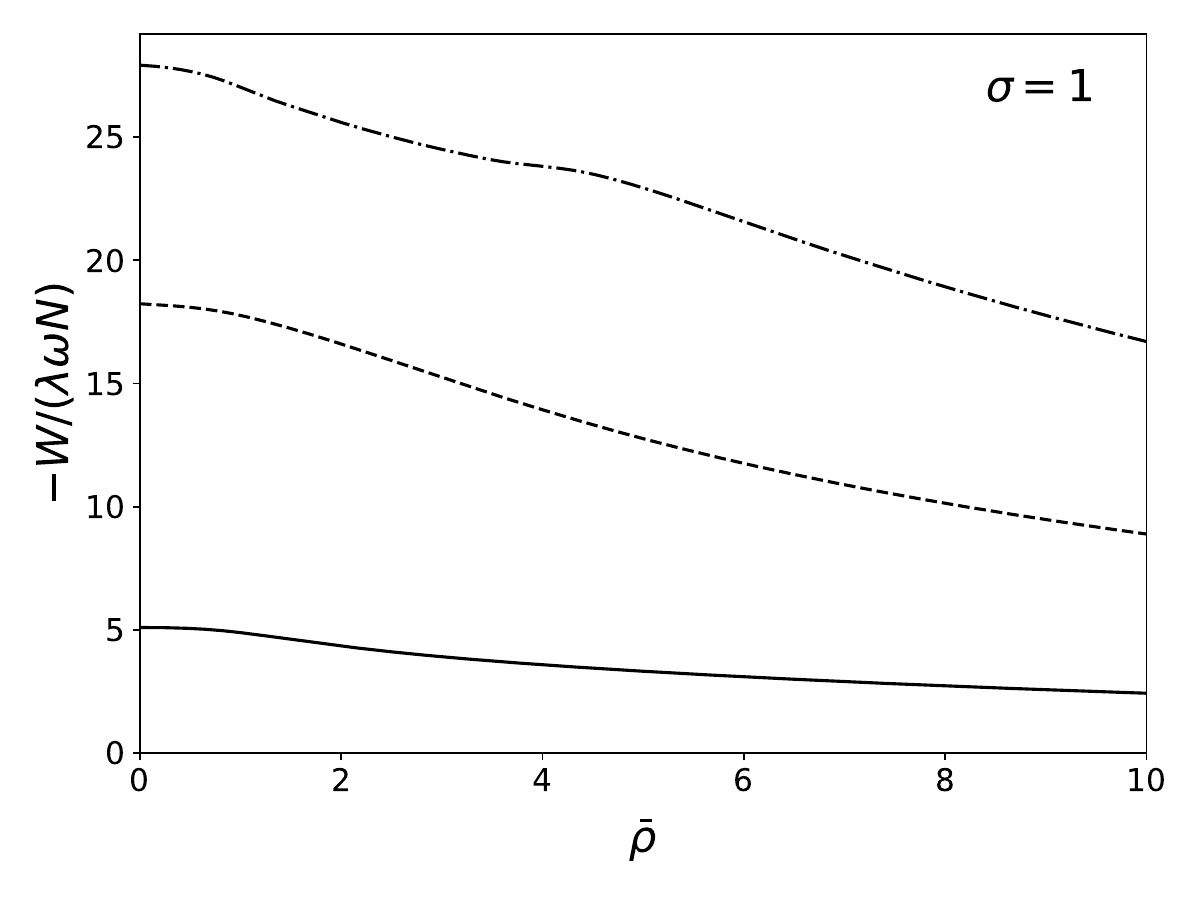}
	\includegraphics[width=0.47\linewidth]{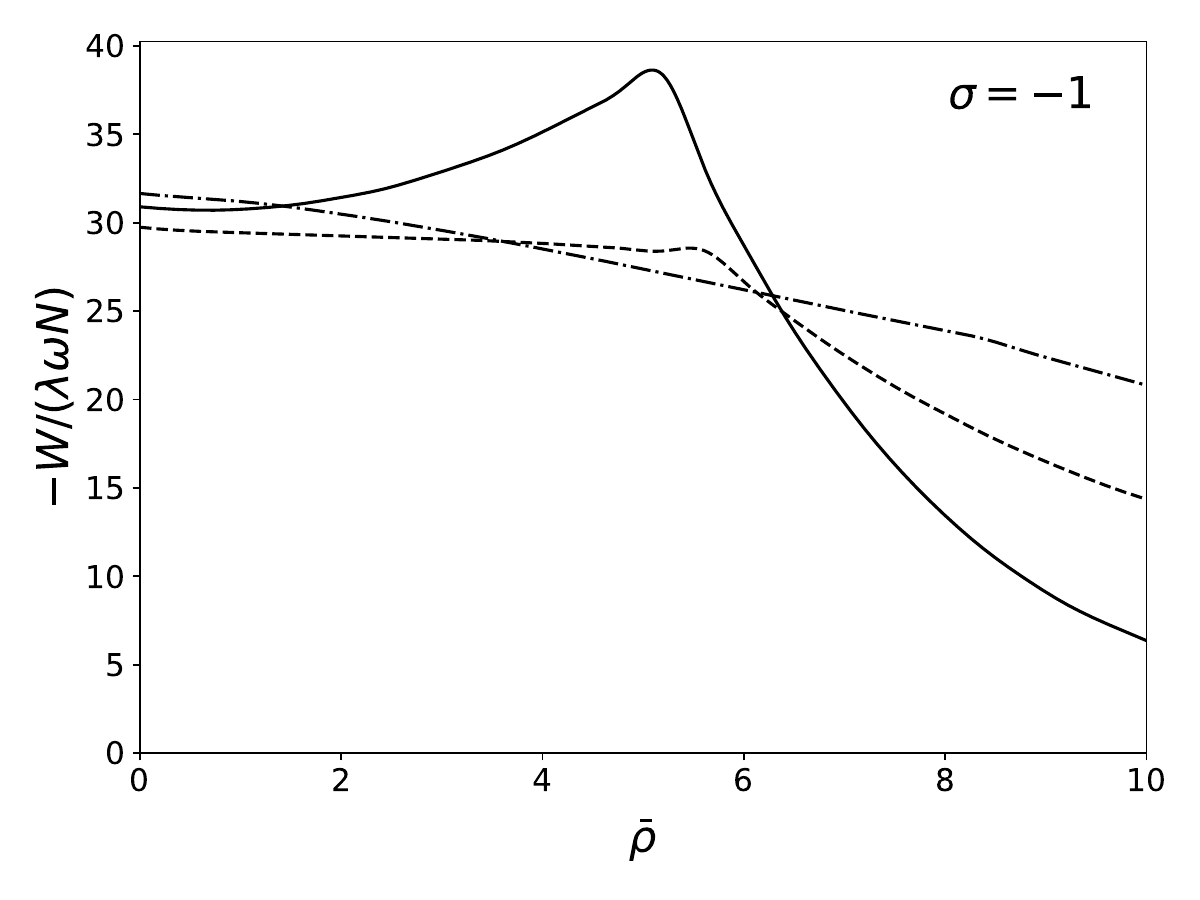}
	\caption{Dependence of $W/(\lambda\omega N)$ on $\bar \rho$ for $\beta=0.6$, $\sigma=\pm 1$, $b_{\max}=100$, $\theta=\pi/8$ (solid line), $\theta=\pi/4$ (dashed line) and $\theta=3\pi/8$ (dash-dotted line).}
	\label{Wfig}
\end{figure}

Fig. \ref{Fyztheta} shows the dependence of $g_y(b)$ and $g_z(b)$ on $\log b$ 
for $\bar\rho =5$, $\beta=0.6$, $\sigma=\pm 1$, $b_{\max}=100$, and several values of
$\theta$. It is evident that for all parameter values, $g_z>0$, while $g_y$ can change sign, and its dependence on $\ln b$ strongly depends on the angle $\theta$.
As before, taking into account nonperturbative contributions is very important, especially for the case $\sigma=-1$. As already mentioned, the friction force leads to a decrease in $\beta$, i.e., to attenuation of the ion beam, see \eqref{W}. In this case, $g_z>0$ and $g_y<0$ lead to a decrease in $\beta_z$ and $\beta_y$, respectively.
However, $g_y>0$ leads to an increase in $\beta_y$. As a result, $\beta$ decreases, and the direction of $\bm\beta$ depends on time in a highly non-trivial way.
\clearpage
Fig. \ref {totFyztheta} shows the dependence of ${\cal F}_y$ and ${\cal F}_z$ on $\theta$ for $\bar\rho=5$, $\beta=0.6$, $\sigma=\pm 1$, and $b_{\max}=100$. By symmetry, $F_y=0$ for $\theta=0$, $\sigma=\pm 1$ and any $\beta$. For $\theta=0$ and $\sigma=1$, we have $F_z=0$ since there is no reflection, and $F_z\neq 0$ for $\sigma=-1$. For $\theta=\pi/2$, we have $F_z=0$ by symmetry, and $F_y\neq 0$.

It is interesting to discuss the dependence of $W$ (the rate of ion energy loss) on $\bar\rho$. Fig.~\ref {Wfig} shows the  function $W$ in units of $\lambda\omega N$ for $\sigma=\pm 1$ in the special case of $\beta=0.6$, $b_{\max}=100$ at various angles $\theta$.

For $\sigma=1$, the function $|W|$  decreases monotonically with increasing $\bar\rho$, while for a fixed $\bar\rho$ this function increases with increasing  $\theta$. Recall that for a fixed magnetic field, the factor $\lambda\omega N$ is constant, and $\bar\rho$ varies due to changes in the transverse temperature of electrons in the beam. For $\sigma=-1$, the behavior of  $|W|$  is more complicated. For small $\bar\rho$, the function $|W|$ depends weakly on $\theta$, in contrast to the case of large $\bar\rho$. Moreover, for small $\theta$, the function $|W|$ has a pronounced maximum at $\bar\rho\sim 2/\beta^2$. From Fig.~\ref{Wfig} it is evident that cooling of positively charged ions  by a positron beam is significantly more efficient than by an electron beam. However, the technical implementation of this idea can be quite nontrivial.

\begin{figure}[h!]
	\centering
	\includegraphics[width=0.47\linewidth]{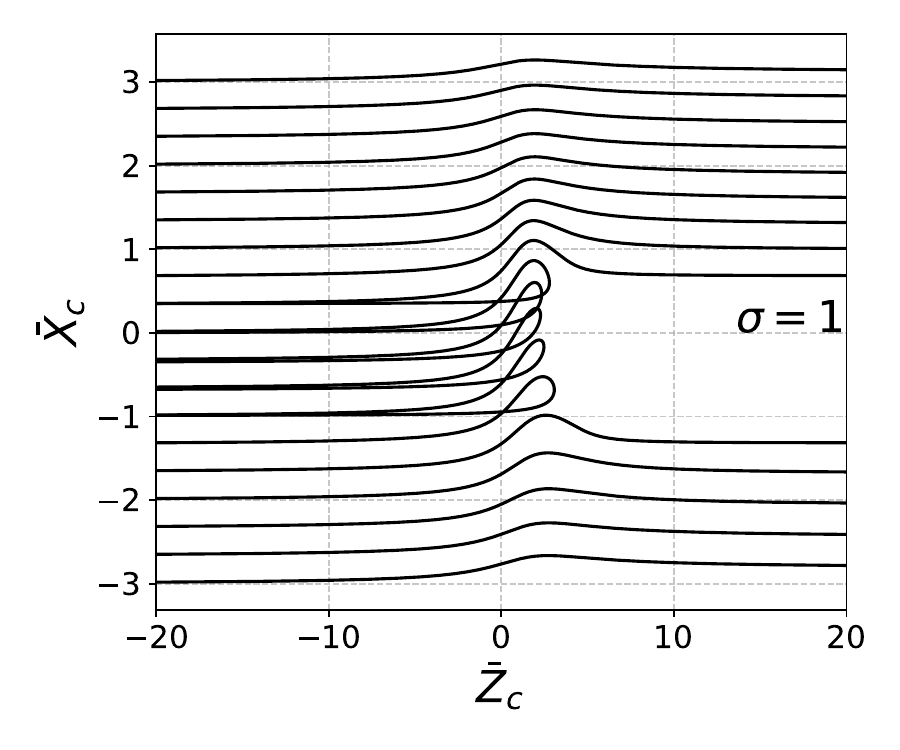}
	\includegraphics[width=0.47\linewidth]{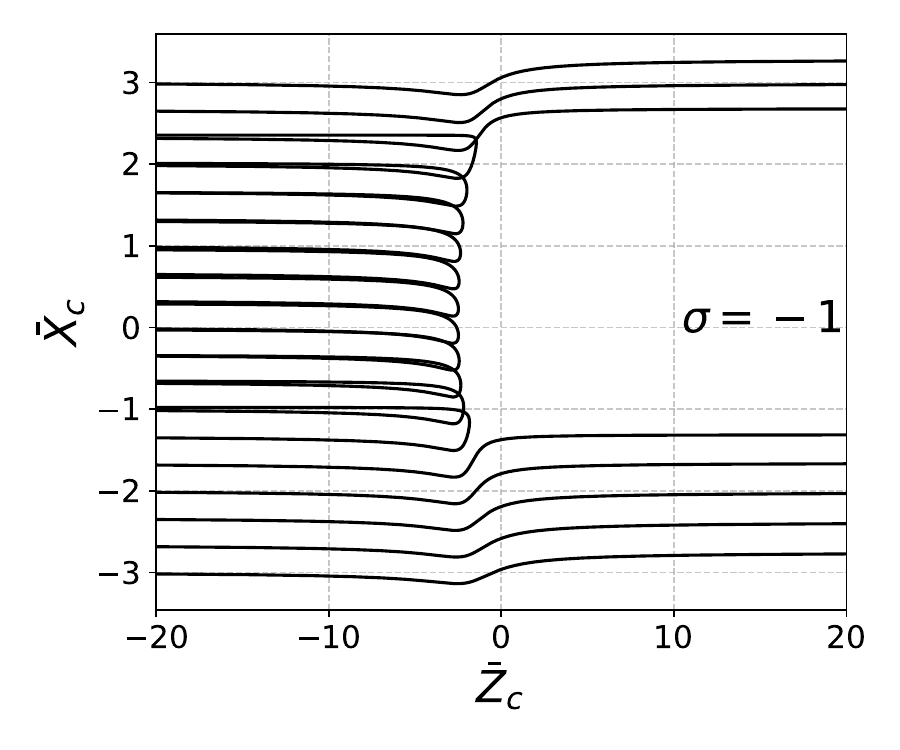}
	\caption{Trajectories of quasiparticles  for $\bar \rho=0$, $\beta=1$, $\theta=\pi/4$, $\sigma=\pm1$, $b_{\max}=100$, $b_y=0$, and various $b_x$.}
	\label{trajectory}
\end{figure}

The above examples of non-trivial dependence of the friction force components on the parameters $\beta$, $\theta$ and $\bar\rho$ in the case of both positively and negatively charged ions are explained by the very unusual trajectories of the quasiparticles, see Fig.\ref{trajectory}.

It is evident that for $\sigma=1$ the region of reflection from the Coulomb center is smaller than in the case of $\sigma=-1$.

\section{Conclusion}
Using the random phase method, we  have reduced the very complex problem of the interaction of an electron beam in a strong magnetic field with ions to the significantly simpler problem of the quasiparticle-ion interaction. As a result, the problem of motion in six-dimensional phase space is reduced to motion in four dimensions. A corresponding Hamiltonian is constructed, which allows calculations of the friction force beyond the framework of perturbation theory. In dimensionless units, quasiparticles in the ion's rest frame are characterized by the parameters $\bm\beta$ and $\bar\rho$. The dependence of the friction force on these parameters is analyzed. It is shown that an account for the effects not described by perturbation theory is fundamentally important when calculating the friction force. This is especially valid for the case of negatively charged ions. The obtained results are important for the design of facilities using the electron cooling.

\section*{Acknowledgments}
We express our gratitude to V.V.~Parkhomchuk, P.V.~Logachev, V.B.~Reva, and V.A.~Vostrikov for helpful discussions.


\begin{thebibliography}{99}
	\bibitem{Budker1967}
	G.I. Budker, Atom. Energ. {\bf 22}, 346 (1967) [Sov. Atom. Energy \textbf {22}, 438 (1967)].
	
	\bibitem{PS1991}
	V.V. Parkhomchuk and A.N. Skrinsky, Rep. Prog. Phys.\textbf{54}, 919 (1991). 
	
	\bibitem{UFN2000}
	V.V. Parkhomchuk, A.N. Skrinsky, Uspekhi Fizicheskikh Nauk \textbf{ 170}, 473 (2000) [Physics-Uspekhi \textbf{43}, 433 (2000)].
	
	\bibitem{UFN2018}
	N.S. Dikansky, I. N. Meshkov, V.V. Parkhomchuk, A.N. Skrinsky, Uspekhi Fizicheskikh Nauk \textbf {188}, 481 (2018) [Physics-Uspekhi \textbf {61}, 424 (2018)].
	
	\bibitem{UFN2025}
	M.I. Bryzgunov, A.V. Bubley, N.S. Dikansky, N.S. Kremnev, V.A. Lebedev, I.N. Meshkov, A.N. Skrinsky, B.N. Sukhina, V.V. Parkhomchuk, D.V. Pestrikov, V.B. Reva,
	Uspekhi Fizicheskikh Nauk \textbf {195}, 101 (2025) [Physics-Uspekhi \textbf{ 68}, 94 (2025)].
	
	\bibitem{DS1978}
	Ya.S. Derbenev , A.N. Skrinsky,  Part. Accel. \textbf {8}, 235 (1978).
	
	\bibitem{Par1984}
	V.V. Parkhomchuk, {\it Physics of Fast Electron Cooling}, Proc. of Workshop on Electron Cooling and Related Applications, Karlsruhe (1984).
	
	\bibitem{Dik1988}
	N.S. Dikansky, V.I. Kudelainen, V.A. Lebedev, I.N. Meshkov, V.V. Parkhomchuk, A.A. Seryj, A.N. Skrinsky, and B.N. Sukhina,  {\it Ultimate possibilities of electron cooling}, Preprint Novosibirsk, INP 88-61 (1988).
	
	\bibitem{LP1980} 
	E.M. Lifshitz and L.P.Pitaevskii, 
	{\it Statistical Physics, Part 2: Theory of the Condensed State} (Pergamon Press, Oxford, 1980).
	
	\bibitem{Meshkov1994}
	I.N.~Meshkov, Phys. Part. Nucl. \textbf{25}, 631 (1994).
	
	\bibitem{ESR1997}
	T.~Winkler, K.~Beckert, F.~Bosch, H.~Eickhoff, B.~Franzke, F.~Nolden, H.~Reich, B.~ Schlitt, and M.~Steck,	Nuclear Physics A \textbf{626}, 485 (1997).

	
\end{thebibliography}
\end{document}